\documentclass[12pt]{article}
\usepackage{alpha}
\usepackage{macros_alpha_ss}
\usepackage{amsmath}
\usepackage{tabularx}
\usepackage{booktabs}
\usepackage[]{appendix}
\usepackage[utf8]{inputenc}
\usepackage{graphicx,amsmath,mathtools,dsfont}
\usepackage{physics}
\usepackage{bigints}

\usepackage{comment}

\newcommand{\Figc}[1]{Figure~}
\newcommand{\Figsc}[1]{Figures~}

\newcommand{\eq}{eq.~}
\newcommand{\Eq}{Eq.~}

\newcommand{\Eqs}{Eqs.~}

\newcommand{\Fig}{Figure~}

\newcommand{\sect}{Sect.~}
\newcommand{\Sec}{Sec.~}

\newcommand{\App}{Appendix~}

\newcommand{\Tab}{Table~}

\newcommand{\Ref}{Ref.~}
\newcommand{\Refs}{Refs.~}

\newcommand{\rmO}{\mathrm{O}}
\newcommand{\SU}[1]{$\mathrm{SU}(#1)$}

\newcommand{\Nc}{N}

\newcommand{\Obs}{\mathcal{O}}

\newcommand{\chidof}{\chi^2/\mathrm{dof}}

\newcommand{\Nmeas}{N_{\mathrm{meas}}}

\newcommand{\Eym}{E(x)}
\newcommand{\Eymt}{E(t)}

\newcommand{\msbar}{\overline{\mathrm{MS}}}
\newcommand{\lat}{\mathrm{lat}}
\newcommand{\GFcoup}{\bar{\lambda}_{\mathrm{GF}}}
\newcommand{\di}{\mathrm{d}}
\newcommand{\GW}{G_W}
\newcommand{\hGW}{\hat{G}_W}
\newcommand{\hW}{\hat{W}}
\newcommand{\HW}{H_W}
\newcommand{\HE}{H_E}
\newcommand{\ats}{a^2/t_0}

\newcommand{\fm}{\mathrm{fm}}

\newcommand{\simas}[1]{\raisebox{-.1ex}{
            $\stackrel{\small{#1}}{\sim}$}}

\newcommand{\tHooft}{'t~Hooft }

\author[desy,hu,epn,espol]{Miguel Garc\'ia Vera}
\author[desy,hu]{Rainer Sommer}

\address[desy]{John von Neumann Institute for Computing (NIC), DESY, \\
    Platanenallee 6, D-15738 Zeuthen, Germany}

\address[hu]{
  Insitut f\"ur Physik, Humboldt Universit\"at zu Berlin,\\
  Newtonstr. 15, D-12489 Berlin, Germany}

\address[epn]{
  Departamento de F\'isica, Escuela Polit\'ecnica Nacional, \\
  Ladr\'on de Guevara, E11-253 Quito, Ecuador
}

\address[espol]{
Departamento de F\'isica, Escuela Superior Polit\'ecnica del Litoral, \\
Campus Gustavo Galindo Km. 30.5 V\'ia Perimetral, Guayaquil, Ecuador
}

\preprintno{DESY 18-075}

\begin{document}

\title{Large $\Nc$ scaling and factorization in \SU{\Nc} Yang-Mills gauge theory}

\begin{abstract}

The large $\Nc$ limit of \SU{\Nc} gauge theories is well understood
in perturbation theory. Also non-perturbative lattice studies have 
yielded important positive evidence that 't~Hooft's predictions are valid.
We go far beyond
the statistical and systematic precision of previous studies
by making use of the Yang-Mills gradient flow and detailed Monte Carlo simulations of  \SU{\Nc} pure gauge theories in 4 dimensions. 
With results for $\Nc=3,4,5,6,8$
we study the limit and the approach to it. We pay particular 
attention to observables which test the expected factorization in the 
large $\Nc$ limit. The investigations are carried out both 
in the continuum limit and at finite lattice spacing. 
Large $\Nc$ scaling is verified non-perturbatively and with high precision; in 
particular, factorization is confirmed. 
For quantities which only probe distances 
below the typical confinement length scale, 
the coefficients of the $1/\Nc$ expansion are of $\rmO(1)$, 
but we found that large (smoothed) Wilson loops
have rather large $\rmO(1/\Nc^2)$ corrections. The exact size
of such corrections does, of course,
also depend on what is kept fixed when the limit is taken.
\end{abstract}

\maketitle

\section{Introduction}

An interesting approach to study quantum chromodynamics (QCD) is to consider 
the order of the gauge group $\Nc$ as a free parameter. As shown by 
\tHooft\cite{'tHooft:1973jz}, by taking the large 
$\Nc$ limit of the perturbative weak coupling 
expansion, the theory simplifies in many ways, and in fact one 
can treat theories at finite $\Nc$ as corrections in the ``small'' 
parameter $1/\Nc$. Moreover, the large $\Nc$ expansion predicts that 
quark loop effects are suppressed by a power of $1/\Nc$, so that the 
weak coupling expansion of large $\Nc$ QCD is dominated by planar 
diagrams with purely gluonic internal loops. All of these rather 
remarkable properties of large $\Nc$ QCD, make it an interesting theory 
to study not only from the theoretical perspective, but also from a 
practical point of view, as results for real world QCD could be 
obtained by considering corrections to the $\Nc = \infty$ theory which 
are parametrized by powers of $1/\Nc$.

Although this $1/\Nc$ scaling is obtained perturbatively, lattice
computations provide evidence that it also holds at the 
non-perturbative 
level, both in $D=4$ space-time dimensions~\cite{Ce:2016awn,Donini:2016lwz,DeGrand:2016pur,Lucini:2004my,GonzalezArroyo:2012fx,Lucini:2012wq,Gonzalez-Arroyo:2014dua} 
and in $D=3$~\cite{Teper:1998te, Lucini:2002wg, Bringoltz:2006zg, Liddle:2008kk, Athenodorou:2016ebg}. 
The evidence is usually 
based on complicated observables, where typically one needs 
to project onto ground states by large Euclidean times.
It is then difficult 
to obtain high precision at various $\Nc$ in order to verify 
\tHooft scaling with good 
confidence. Let us stress the fact that the 
validity of the $1/\Nc$ scaling, beyond the weak coupling expansion,  
is not a trivial statement. Hence, it 
is desirable to test it by means of lattice simulations
and with statistically and systematically very precise observables. 

Perturbatively, if one carries on with the \tHooft $1/\Nc$ topological 
expansion, another simplification arises, which
has to do with the property of factorization

\begin{equation}
	\ev{\Obs_1 \Obs_2} = \ev{\Obs_1} \ev{\Obs_2} + \rmO(1/\Nc^2)\, ,
	\label{eq:basic_fact}
\end{equation}
where the $\Obs_i$ are local gauge invariant or Wilson loop 
operators, and the leading correction scales as $1/\Nc^2$ in
the pure gauge theory, which we focus on for the rest of this work. \Eq\eqref{eq:basic_fact}
has several consequences, as it tells us that in the large $\Nc$ limit, 
the dominant part of a correlator is the disconnected one. In 
particular, when $\Obs_1 = \Obs_2$, this means that fluctuations are 
suppressed; and as discussed in \Ref\cite{Yaffe:1981vf}, this fact can 
be put in analogy with the classical limit of a quantum theory, where 
$1/\Nc$ plays the role of $\hbar$. 
Related to this is also the concept 
of the ``master field'', i.e., the idea that the path integral is 
dominated by a single gauge configuration 
(or rather a gauge orbit)~\cite{Witten:1979pi,Coleman:1980nk}. Although 
these ideas triggered hope to find the solution 
of large $\Nc$ QCD, such an analytical solution is still lacking today.
The situation in the Yang-Mills theories is similar in this respect 
to two-dimensional \SU{\Nc}$\times$\SU{\Nc} spin models\cite{Rossi:1996hs}, while 
for O$(N)$ models and CP$(N)$ models the large $N$ limit is solvable 
and one can therefore really carry out the expansion \cite{DIVECCHIA1984478,Campostrini:1993gya,Campostrini:1993fr}.

One more aspect where \Eq\eqref{eq:basic_fact} plays a crucial role 
has to do with the idea of volume independence, which starting from the 
work of the authors in \Ref\cite{Eguchi:1982nm}, has been used in the 
lattice formulation to study the large $\Nc$ limit of the Yang-Mills 
theory by performing simulations in small spacetime 
volumes~\cite{Kiskis:2003rd}, and even 
in single site lattices, provided a clever choice of boundary 
conditions~\cite{GonzalezArroyo:1982hz,GonzalezArroyo:1982ub,
GonzalezArroyo:2010ss} is made. Clearly, the possibility to compute 
observables on a single site lattice makes simulations of \SU{\Nc} 
Yang-Mills theory at large $\Nc$ more accessible, as there is a significant
compensation of the extra cost for 
increasing $\Nc$ by the much smaller number of lattice sites.

The above indicates that factorization is not only relevant in the 
theoretical context, but also on the practical level, as it is a 
requirement for the single site lattice simulations to be valid. To be 
more precise, the equivalence between the single site and the infinite 
volume theory is argued for on the basis of the Makenko-Migdal loop 
equations~\cite{Makeenko:1979pb} on the lattice. As originally shown in 
\Ref\cite{Eguchi:1982nm}, the loop equations in both theories are 
equivalent, provided that the product of the expectation value of the 
Wilson loops factorize as stated in \Eq\eqref{eq:basic_fact}. 
Additionally, we would like to point 
out that important physics is contained in the corrections to 
factorization. The most obvious one is that glueball masses are 
obtained from the connected correlation functions of Wilson loops. 

The previous discussion motivates the search for
a non-perturbative proof, beyond the realm of 
weak coupling perturbation theory.
 Several authors have investigated
factorization beyond perturbation theory~\cite{Chatterjee:1984pa,
Makeenko:1999hq,Chatterjee:2015exa,Jafarov:2016ngk}, but no
verification has been carried through using the non-perturbative 
framework provided by numerical simulations of 
lattice gauge theories.
We here fill this gap.
In addition, the observables that we consider make it possible
to study the large $\Nc$ scaling up to very high precision, and hence 
address the important issue of the size of the corrections to $\Nc=\infty$.

This paper is organized as follows, in \Sec 2 we present the 
observables that are used both to check the large $\Nc$ scaling, as well
as factorization. In \Sec 3 we discuss different ways of defining the 
large $\Nc$ limit and in particular the two choices we made for our investigation.
In \Sec 4 we describe the ensembles and lattice 
parameters used for the simulations and in \Sec 5 we present our 
results, both at finite lattice spacing, and in the continuum limit. We
finish with a short summary of the results.

\section{Observables}\label{sec:observables}

The basic observables we consider are the Yang-Mills action density
$E(t)$ at positive flow time \cite{Luscher:2011bx} (defined below)
and rectangular Wilson loop operators
\begin{equation}
	W_C = \frac{1}{\Nc} \tr P \left\{ \exp\left( \oint_C A_{\mu}(x) \, \di x^{\mu} \right)  \right\} \, ,
	\label{eq:Wloop_cont}
\end{equation}
where $C$ is a closed rectangular path in space-time, and $P$ denotes 
the path ordering operator. The normalization factor $1/\Nc$ is included
in the definition of $W_C$ in order to have a finite large $\Nc$ limit ---
already at tree-level.
Wilson loops have singularities which have to
be removed before the continuum limit can be taken.
In particular, 
for our square Wilson loops, one must remove not just the 
``perimeter'' divergences but also 
``corner'' divergences~\cite{Dotsenko:1979wb,Brandt:1981kf,Dorn:1986dt}. 
One way 
to proceed is to consider Creutz 
ratios~\cite{Creutz:1980hb}, which however, for loops of large 
size in lattice units, suffer from small signal to noise ratios. 
As this would compromise our desire for a precision test,
we work instead  with smooth Wilson loops. 
The
smoothing is provided by the Yang-Mills gradient 
flow~\cite{Narayanan:2006rf,Luscher:2010iy}. It evolves the gauge 
fields $A_{\mu}(x)$ according to the flow equation
\begin{align}
	\partial_t B_{\mu}(t,x) &= D_{\nu} G_{\nu\mu}(t,x)\, , 
	\quad B_{\mu}(0,x) = A_{\mu}(x) \,  \\\nonumber
	G_{\mu\nu}(t,x)&= \partial_{\mu}B_{\nu}(t,x) - 
	\partial_{\nu}B_{\mu}(t,x) - 
	\left[ B_{\mu}(t,x), B_{\nu}(t,x) \right] \, ,
	\label{eq:GFlow continuum}
\end{align}
where the dimension two parameter $t$ is known as the flow time. The 
loops at positive flow time are
then simply 
\begin{equation}
	W_C(t) = \frac{1}{\Nc} \tr P \left\{ \exp\left( \oint_C B_{\mu}(t,x) \, \di x^{\mu} \right)  \right\} \, .
	\label{eq:Wloop_cont}
\end{equation}
Choosing $8t$ to be of a typical QCD size, say of the order of the
square of the inverse string tension, they benefit from small statistical errors even for large loops~\cite{Gonzalez-Arroyo:2014qza}. In particular, their variance remains finite in the 
continuum limit.
That property is a particular manifestation of the
most important feature of observables which are built from the smoothed gauge 
fields $B_{\mu}(t,x)$: they are renormalized operators at 
positive flow time $t$~\cite{Luscher:2011bx,Lohmayer:2012ue}. 
In other words, there is no renormalization scheme or scale 
dependence beyond $t$ and the continuum limit is unambiguous and well defined.
Even the action density
\begin{equation}
	\Eym= -\frac{1}{2} 
	\tr \left\lbrace G_{\mu\nu}(x) G_{\mu\nu}(x) \right\rbrace\, ,
	\label{eq:Eymdef}
\end{equation}
is finite. It will be one of our observables.

\subsection{The gradient flow coupling at large $\Nc$}

The gradient flow can also be used to define a renormalized 
coupling~\cite{Luscher:2010iy}. Using the perturbative expansion of the 
Yang-Mills energy density at positive flow time $\ev{\Eymt}$, one has 
that  

\begin{equation}
	\GFcoup(\mu) = \frac{128 \pi^2}{3}\left( \frac{N}{N^2-1} \right)t^2 \ev{\Eymt}\Big\rvert_{\mu=1/\sqrt{8t}} 
	= \bar{\lambda}_{\msbar}(\mu) \left[ 1 + c_1 \bar{\lambda}_{\msbar}(\mu) + \cdots \right]\, ,
\label{eq:Gflow_coupling}
\end{equation}
where 
$\bar{\lambda}_{\msbar}(\mu) = N\bar{g}^2_{\msbar}(\mu)$
is the \tHooft coupling at the scale $\mu = 1/\sqrt{8t}$ and
$c_1 = \frac{1}{16 \pi^2} \left( \frac{11}{3}\gamma_E + \frac{52}{9} 
- 3 \ln 3 \right)$ is $\Nc$ independent. With this definition, we can 
then define a scale by setting the renormalized coupling 
$\GFcoup$ to a given value. A convenient choice for \SU{3} is the 
reference scale $t_0$~\cite{Luscher:2010iy}, which 
corresponds to a
value of the coupling such that $t^2\ev{\Eymt}|_{t=t_0} = 0.3$. This 
particular choice can be generalized to \SU{\Nc} if the right hand side 
is modified so that it has the correct scaling with $\Nc$. Clearly, we 
also want the definition to remain what it is for $\Nc=3$.
Thus, 
\Eq\eqref{eq:Gflow_coupling}, suggests to define $t_0$ implicitly by the equation~\cite{Ce:2016awn}
\begin{equation}
	t^2\ev{\Eymt}\rvert_{t=t_0} = 0.1125 \, \frac{\Nc^2 -1}{\Nc}\, ,
	\label{eq:t0SUN}
\end{equation}
for all $\Nc$. 

\subsection{Smooth Wilson loops}\label{sec:smoothloops}

The favourable properties of smooth
Wilson loops have already been exploited in the literature, as for 
example to estimate the string tension at small values of $t$ in
\Refs\cite{GonzalezArroyo:2012fx,Lohmayer:2012ue}, or to study the large
$\Nc$ phase transition in the eigenvalue spectrum of the Wilson loop 
matrices~\cite{Lohmayer:2011nq}. For our purpose, the limit of small $t$
is not required, as the smooth loops are used to test factorization 
and the large $\Nc$ limit for 
well defined renormalized observables, regardless of their relation to 
the operators at $t=0$.

\begin{figure}
	\centering
	\includegraphics[width=0.4\textwidth]{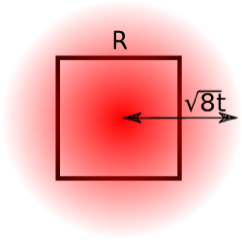}
\caption{Schematic representation of a smooth Wilson loop operator. 
	The size of the loop is chosen such that it has the same 
length as the smoothing radius, i.e. $R=\sqrt{8t}$. }
	\label{fig:smooth_loop}
\end{figure}

In the end, we study the large $\Nc$ limit of square Wilson 
loops, i.e. for loops where the path $C$ in 
\Eq\eqref{eq:Wloop_cont} is given by a square of size $R\times R$. In order to 
take the large $\Nc$ limit, the loops are matched at different $\Nc$ 
relating their size to the scale $t_0$ introduced in the previous 
section. More precisely, the large $\Nc$ and continuum limits are taken 
for loops of size $R_c = \sqrt{8 c t_0}$ (see \Fig\ref{fig:smooth_loop}), 
where the smoothing parameter $t = c t_0$, and $c$ is a constant 
parameter.

To be more precise, let us denote a square loop with one of its corners 
at the spacetime point $(x_0,\vec{x})$ and
extending only in space as $W(t,x_0,\vec{x},R)$. Its expectation value
\begin{equation}
	W(c)= \left\langle  W(t,x_0,\vec{x},R_c) \right\rangle\, 
	\quad \text{with }
	\quad t=ct_0\,, R_c = \sqrt{8 c t_0}\,,\; 
	\label{eq:Wdef}
\end{equation}
is independent of the position $\vec{x}$ due to translation invariance
and only depends on the parameter $c$. In our notation we separate 
time and space-coordinates, as we will later use lattices with 
different boundary conditions in time (open b.c.) and space (periodic b.c.).
While the independence on $\vec{x}$ is exact, $x_0$ has to be sufficiently
far away from the boundary for  \Eq\eqref{eq:Wdef} to hold.

Similarly, we define 

\begin{equation}
	W^{\mathrm{sq}}(c)= \left\langle  W^2(t,x_0,\vec{x},R_c) \right\rangle\, 
	\quad \text{with }
	\quad t=ct_0\,, R_c = \sqrt{8 c t_0}\,,\; 
	\label{eq:W2def}
\end{equation}
which corresponds to the expectation value of the product of a Wilson loop with 
itself.

\subsection{Observables to test factorization}\label{sec:factobs}

In order to investigate the property of factorization from 
\Eq\eqref{eq:basic_fact}, we define several observables based on the 
Yang-Mills action density and the smooth Wilson loops at positive flow 
time. They are constructed such that factorization implies
that they vanish as 
$\Nc \to \infty$. First we consider the simplest case of the observable 
$\GW$ defined in terms of the smooth Wilson loops as
\begin{equation}
	\GW(c) = \frac{W^{\mathrm{sq}}(c) - W^2(c)}{W^2(c)} \, .
	\label{eq:Gobservable}
\end{equation}

Then, we consider observables built from the space integral of 
the smooth Wilson loops and the Yang-Mills action density. We 
define\footnote{In the lattice discretisation, one just needs
to replace $\int \di^3x \to a^3 \sum_{\vec{x}}$}
\begin{equation}
	H_{\Obs}(c) = \left( \frac{1}{t_0^{3/2}} \right) 
	\frac{ {\bigint} \di^3\vec{x} \, \left[ \ev{\Obs(ct_0,x_0,\vec{x})\Obs(ct_0,x_0,\vec{0})} - \ev{\Obs(ct_0,x_0,\vec{x})}^2 \right]}{\ev{\Obs(ct_0,0,\vec{x})}^2} \, ,
	\label{eq:Hobservable}
\end{equation}
with the factor $1/t_0^{3/2}$ rendering $H_{\Obs}(c)$ 
dimensionless, and where $\Obs$ is either a smooth Wilson loop, or 
the Yang-Mills action density. Notice that $H$ is a type of 
susceptibility, as we are integrating over the contributions from the 
correlation function of $\Obs$ at different distances. The integration 
does not extend over $x_0$ due to our choice of boundary conditions
and $x_0$ is again supposed to be far away from the time-boundaries.
In comparison to the simple 
observable $\GW$, this probes longer distances, but introduces also
more noise and affects the statistical 
errors in the measurements. Nonetheless, as will be shown in 
\Sec\ref{sec:results}, the statistical precision that can be achieved 
for $H_{\Obs}$ remains good.
In particular, we will consider $\HE(c)$, defined by 
inserting
\begin{equation}
	\Obs(t,x_0,\vec{x}) = 
	E(t,x_0,\vec{x}) \, ,
	\label{eq:HEobservable}
\end{equation}
into \Eq\eqref{eq:Hobservable} and $\HW$ by
\begin{equation}
	\Obs(t,x_0,\vec{x}) = W(t,x_0,\vec{x},R_c) \, .
	\label{eq:HWobservable}
\end{equation}
We remind the reader of our choice $R_c = \sqrt{8 c t_0}$.

\Eq\eqref{eq:basic_fact} means

\begin{equation}
	\HE \simas{\Nc\to\infty}\, 1/N^2\,,\;\;  	
	G_W \simas{\Nc\to\infty}\, 1/N^2\,,\;\;
	H_W \simas{\Nc\to\infty}\, 1/N^2\,.
	\label{eq:limFGH}
\end{equation}

\subsection{Finite volume}
For a numerical test, we need to choose a finite volume.
We chose our parameters such that $L/\sqrt{8t_0}\approx 3.3$.
Table~\ref{tab:fact_ensembles} shows the actual values used in our 
simulations. Since $L$ is thus approximately constant, it 
is omitted as an argument of the observables. We note that the large
$\Nc$ limit and factorization can be tested in infinite or in finite volume. 
To be on the safe side, we chose the latter, even though we are not far from the 
infinite volume limit for most observables.

\section{Defining the approach to the large $N$ limit}\label{sec:limit}

The complete definition of a quantum field theory involves a regularization
(here Wilson's lattice theory)
as well as a non-trivial renormalisation before 
the regulator can be removed. Although this is usually 
not discussed, quantitative statements about the approach 
to the large $\Nc$ limit,
such as the ones we are seeking here, do depend on the 
renormalisation scheme if the renormalisation scheme
defines which quantity is held fixed as we take $\Nc\to\infty$.

While the $\rmO(1/\Nc^2)$ corrections depend on these 
details, the true limit is expected to be unique in the following sense.
It is independent of the scheme, as long
as the \tHooft coupling $\bar{\lambda}_{s}(\mu) = \Nc\bar{g}^2_{s}(\mu)$
in any scheme is kept fixed as one takes the limit.  
This statement becomes most transparent when we replace couplings 
by the associated $\Lambda$-parameters,
\begin{equation}
	\Lambda_s \, =  \lim_{\mu\to\infty} \,
	\mu\, 
	\left( \frac{48\pi^2}{11 \lambda_s(\mu)} \right)^{51/121} 
	\exp\left( -\frac{1}{b_0\lambda_s(\mu)} \right) \,, \quad b_0 = \frac{11}{24\pi^2} \,
	.
	\label{eq:lambda_s}
\end{equation}
Now any renormalization group invariant 
quantity $\Obs$ of 
mass dimension $n$, has a large $\Nc$ limit
\begin{equation}
    \lim_{\Nc\to\infty} \frac{\Obs}{\Lambda_s^n} 
    = \lim_{\Nc\to\infty} r^n(\Nc) \frac{\Obs}{\Lambda_{s'}^n} 
     = r_\infty^n\,  \lim_{\Nc\to\infty} \frac{\Obs}{\Lambda_{s'}^n}\,,
	\label{eq:uniqueness}
\end{equation}
where 
\begin{eqnarray}  
  r(\Nc) &=& \exp(c_{ss'}(\Nc)/b_0) \,,
  \\
  \lambda_{s'} &=& \lambda_{s} +c_{ss'}(\Nc)\lambda_{s}^2 + \rmO(\lambda_{s}^3)\,.
\end{eqnarray}
and
\begin{equation}  
  r_\infty = \exp(\lim_{\Nc\to\infty} c_{ss'}(\Nc)/b_0) \,.
\end{equation}

Examples for $n=1$ are glueball masses and 
$t_0$ defined above is a RGI scale  with $n=-2$. When the observable 
$\Obs$ depends on  external momenta or coordinates, they have to be fixed
in units of $\Lambda$ in a specified scheme, e.g. $\Lambda_{\msbar}$,
when taking $\Nc\to\infty$. 

Due to the existence of the limit \eq\eqref{eq:uniqueness}, we may also 
scale distances with respect to any one particular reference scale 
(choice of $\Obs$). In our numerical work we have chosen
$t_0$, \eq\eqref{eq:t0SUN}, because of its high precision.

The preceding discussion is about the continuum theory. It thus saliently assumes that first we take the continuum limit at finite $\Nc$ and
then we perform $\Nc\to\infty$. However, we may also 
proceed in the opposite order: first take the large $\Nc$ limit at fixed 
lattice spacing and then send the lattice spacing to zero.\footnote{
Numerically this is of interest because at not-so-small lattice 
spacing the first step can easily be investigated with a larger 
range in $\Nc$. Even more,
as shown in the next section, the results at finite 
lattice spacing can be obtained with higher precision as only an 
\emph{interpolation} to a common lattice spacing for all $\Nc$  
is needed, and thus the statistical and systematic errors 
are greatly reduced when compared to the
results of a continuum limit \emph{extrapolation}.} 
Let us briefly 
discuss that this order of limits is indeed the same as above; the limits are interchangeable.

\subsection{Large $\Nc$ limit at fixed lattice spacing}

The existence of the large $\Nc$ limit at fixed finite lattice spacing 
is expected due to the following
consideration.
We start from the Lambda-parameter,
$\Lambda_\lat$ in the lattice minimal subtraction scheme, 
which satisfies \eq\eqref{eq:lambda_s} with $\mu=1/a$ and 
$\lambda_\lat(\mu)=\lambda_0$ in terms of the lattice spacing, $a$, and
the bare coupling, $\lambda_0=\Nc g_0^2$. In fact, having a specific scheme, the lat-scheme, we can 
give the more detailed formula,
\begin{equation}
	a \Lambda_\lat \, =  \,\, \left( \frac{48\pi^2}{11 \lambda_0} \right)^{51/121} 
	\exp\left( -\frac{24\pi^2}{11 \lambda_0} \right)\left( 1 + c_1(\Nc) \lambda_0 
	+\rmO(\lambda_0^2)\right) \, ,
	\label{eq:lambda_lat}
\end{equation}
where $c_1 = 0.1048 + \rmO(1/\Nc^2)$~\cite{Alles:1996cy}. 
\Eq\eqref{eq:lambda_lat} shows that the large 
$\Nc$ limit can be taken at fixed bare coupling which is
equivalent to fixed lattice spacing $a$. 
{\em Apart from $\rmO(1/\Nc^2)$ terms} in $c_1$ and higher order terms, 
fixed lattice spacing is the same as fixed $\Lambda_\lat$ and
therefore also fixed $\Lambda_s$ in other schemes.
See also an early discussion of $\Lambda_{\msbar}/\Lambda_\lat$ including its $\Nc$
dependence~\cite{Dashen:1980vm}.

In general, taking the large $\Nc$ limit at fixed lattice spacing has to be
followed by the $a\to0$ limit at $\Nc=\infty$. However, when we investigate
factorization, the second step is not expected to be necessary. 
This is because the perturbative proof of factorization holds in the
lattice regularization \cite{tHooft:2002ufq} at finite $a$. If factorization holds 
non-perturbatively we thus also expect \eq\eqref{eq:limFGH} at any fixed $a$. In any case, verifying \eq\eqref{eq:limFGH} at arbitrary finite
lattice spacing implies that it holds in the continuum limit.

Note also that even the large $\Nc$ limit of 
divergent quantities, such as Wilson loops at $t=0$, 
is expected to exist. A high precision numerical test has recently been performed \cite{Gonzalez-Arroyo:2014dua}.

\section{Lattice details}

In this section we give the details of our lattice simulations. 
We simulate the pure gauge theory with $\Nc=3,\, 4,\, 5,\, 6,\, 8$ at several 
lattice spacings. The lattice action is the Wilson gauge action and we use 
open boundary conditions in the time direction~\cite{Luscher:2011kk}.
The simulations are performed using a combination of heatbath and 
overrelaxation local updates using the Cabibbo-Marinari 
strategy~\cite{Cabibbo:1982zn} to refresh the \SU{\Nc} matrices. The 
ratio of overrelaxation to heatbath updates is fixed to $L/(2a)$. 

For convenience, we present the values of the lattice spacing, as well 
as lattice sizes in physical units by assigning a value to $t_0$ such 
that $\sqrt{t_0}=0.166 \, \fm$. This choice is motivated by the result 
in \SU{3} for $\sqrt{8t_0}/r_0 = 0.941(7)$~\cite{Ce:2015qha} and the 
value of the reference scale $r_0 \approx 0.5 \, \fm$~\cite{Sommer:1993ce}.
Notice that this choice is somewhat arbitrary, as apart from the missing
quark loops, for $\Nc \neq 3$ the 
theory cannot be directly identified with Nature. 

The parameters of the simulations are displayed in 
\Tab\ref{tab:fact_ensembles}. The configurations used for the 
measurements are a subset of those reported in \Ref\cite{Ce:2016awn} for 
all ensembles except for those at $\Nc =3, 8$, and for the finest lattice
spacings in the case of $\Nc=4, 5$.  As announced above, all the lattices considered in 
\Tab\ref{tab:fact_ensembles} are of approximately the same spatial size 
$L \approx 1.55 \, \fm$. In addition, we have used two additional 
ensembles with $L \approx 2.35 \, \fm$ at the coarsest lattice spacing 
($a \approx 0.1 \, \fm$) for $\Nc = 4, 5$ in order to check for 
effects due to small variations in the volume. 
Notice that for the ensembles which have been reported 
in \Ref\cite{Ce:2016awn}, we have a very large number of measurements 
for the Yang-Mills action density. 

\begin{table}[tb]
  \centering
  \begin{tabular}{ccccccrrcc}
    \toprule
    {\#run} & {$\Nc$} & {$\beta$}  & {$T/a$} & {$L/a$} & {$a[\fm]$} & {$\Nmeas^\mathrm{W}$} & {$\Nmeas^\mathrm{E}$} & $t_0/a^2$ & $ L/\sqrt{8t_0}$\\
    \midrule
    $A(3)_2$ & 3 & 6.11  & 80 & 20  & 0.078 & 320 & 6720 & $4.5776(15)^{*}$ & 3.3050(5)\\
    $A(3)_3$ & 3 & 6.24  & 96 & 24  & 0.064 & 280 & 280 & $6.783(23)$ & 3.258(6) \\
    $A(3)_4$ & 3 & 6.42  & 96 & 32  & 0.050 & 252 & 252 & $11.19(4)$ & 3.382(6)\\
    \midrule
    $A(4)_1$ & 4 & 10.92 & 64 & 16  & 0.096 & 248 & 17341 & $2.9900(7)^{**}$ & 3.2714(4) \\
    $A(4)_2$ & 4 & 11.14 & 80 & 20  & 0.078 & 300 & 35960 & $4.5207(8)^{**}$ & 3.3257(3) \\
    $A(4)_3$ & 4 & 11.35 & 96 & 24  & 0.065 & 312 & 15460 & $6.4849(16)^{**}$ & 3.3321(4)\\
    $A(4)_4$ & 4 & 11.65 & 96 & 32  & 0.049 & 320 & 640 & $11.55(3)$ & 3.329(4)  \\
    \midrule                                                  
    $A(5)_1$ & 5 & 17.32 & 64 & 16  & 0.095 & 320 & 9871 & $3.0636(7)^{**}$  & 3.2319(4)\\
    $A(5)_2$ & 5 & 17.67 & 80 & 20  & 0.077 & 240 & 21680 & $4.6751(8)^{**}$ & 3.2703(3)\\
    $A(5)_3$ & 5 & 18.01 & 96 & 24  & 0.064 & 248 & 8007 & $6.8151(18)^{**}$ & 3.2504(4)\\
    $A(5)_4$ & 5 & 18.21 & 96 & 32  & 0.049 & 328 & 328 & $11.51(3)$ & 3.334(4)  \\
    \midrule                                                  
    $A(6)_1$ & 6 & 25.15 & 64 & 16  & 0.095 & 320 & 19360 & $3.0824(4)^{**}$ & 3.2220(2) \\
    $A(6)_2$ & 6 & 25.68 & 80 & 20  & 0.076 & 264 & 11392 & $4.8239(9)^{**}$ & 3.2195(3)\\
    $A(6)_3$ & 6 & 26.15 & 96 & 24  & 0.063 & 288 & 6704 & $6.9463(13)^{**}$ & 3.2195(3)\\
    \midrule                                                       
    $A(8)_2$ & 8 & 32.54 & 20 & 80  & 0.076 & 320 & 320 & $4.782(5)$ & 3.2336(17)\\ 
    \bottomrule
  \end{tabular}
  \caption{Parameters of the simulations. For each of the gauge groups 
   \SU{\Nc} we give the inverse lattice coupling $\beta=2\Nc^2/\lambda_0$,
   the dimensions of the lattice, the approximate lattice spacing using 
   $\sqrt{t_0}=0.166 \, \fm$ followed by the number $\Nmeas^\mathrm{W}$ of
   measurements used for the computation of the smooth Wilson loops, and $\Nmeas^\mathrm{E}$ for the action  density, \eq\eqref{eq:Eymdef}. 
   In the second to last column we present the values of 
   $t_0/a^2$:$~^{*}$ taken from \Ref\cite{Vera:2016xpp} and$~^{**}$ 
   taken from \Ref\cite{Ce:2016awn}.
  \label{tab:fact_ensembles}
  }
\end{table}

The flow equations are integrated using a third order Runge-Kutta 
integrator~\cite{Luscher:2010iy} and the observables are measured at 
intervals $\Delta t$ of $t$ of $\Delta t/a^2 \approx 2-3 \times 10^{-2}$. 
Afterwards, they are 
interpolated using a second order polynomial in order to obtain their 
values at arbitrary $t$. The action density is 
defined exactly as in \cite{Luscher:2010iy}, 
using the clover discretization and it is measured from $t=0$ 
up to $t \approx 1.2 \, t_0$. The loops, $W(c)$, are
measured only in the vicinity of $t=ct_0$, with 
$c={1/2,1,9/4}$, and then interpolated to the exact value of $t$. For 
the loops, one has to do an additional 
interpolation to $R_c$, and since their statistical precision is very 
high, one has to be careful with small potential systematic effects.
The details of this interpolation were already presented 
in~\Ref\cite{Vera:2017dxr}.

We end this section with the precise definition of the observables 
introduced in \Eqs\eqref{eq:Wdef}-\eqref{eq:Hobservable} on the 
lattice. First, for the Wilson loops we use translation invariance in 
the form 

\begin{align}\nonumber
W(c) &= 
 \frac{a^4}{\left( T - 2d \right) L^3} \sum_{x_0=d}^{T-d-a} 
 \sum_{\vec{x}} \ev{W(ct_0,x_0,\vec{x},R_c)} \, ,\\
 W^{\mathrm{sq}}(c) &=  
 \frac{a^4}{\left( T - 2d \right) L^3} \sum_{x_0=d}^{T-d-a} 
 \sum_{\vec{x}} \ev{W(ct_0,x_0,\vec{x},R_c)^2} \, ,
\end{align}
where $\ev{\cdot}$ corresponds to the estimator of the true expectation value 
computed on the lattice. 

In order to compute $H_W$, we define

\begin{equation}
W^{\mathrm{sq}}_{\mathrm{int}} (c) = \frac{a}{(T-2d)} \sum_{x_0 = d}^{T-d-a} 
	\ev{ \left( \frac{a^3}{L^3}  
	\sum_{\vec{x}} W(ct_0,x_0,\vec{x},R_c) \right)^2} \, ,
\end{equation}
so that 

\begin{equation}
	\HW(c) = \left( \frac{L^3}{t_0^{3/2}} \right) 
	\frac{W^{\mathrm{sq}}_{\mathrm{int}}(c) - W^2(c)}{W^2(c)} \,  ,
	\label{eq:HWlattice}
\end{equation}
and we proceed in a similar way to define $\HE$ after replacing 
$W(ct_0,x_0,\vec{x},R_c)$ by $t^2 E(ct_0,x_0,\vec{x})\rvert_{t=ct_0}$. 
The parameter $d$ is 
introduced to deal with the systematic effects from the open boundary 
conditions. It is chosen in a similar way as described in 
\Ref\cite{Ce:2016uwy}, so that the effects coming from the boundaries 
are negligible with respect to the statistical error in the bulk.

\section{Results}\label{sec:results}

\subsection{Large $\Nc$ scaling}

In order to test and provide a precise verification of 
$1/\Nc^2$ scaling, we analyse our results for $W(c)$
and for the gradient flow coupling $\GFcoup$. Let us first discuss our 
results for the latter.

\subsubsection{The gradient flow coupling}\label{sec:resgfcoup}

\begin{figure}
	\centering
	\includegraphics[width=0.7\textwidth]{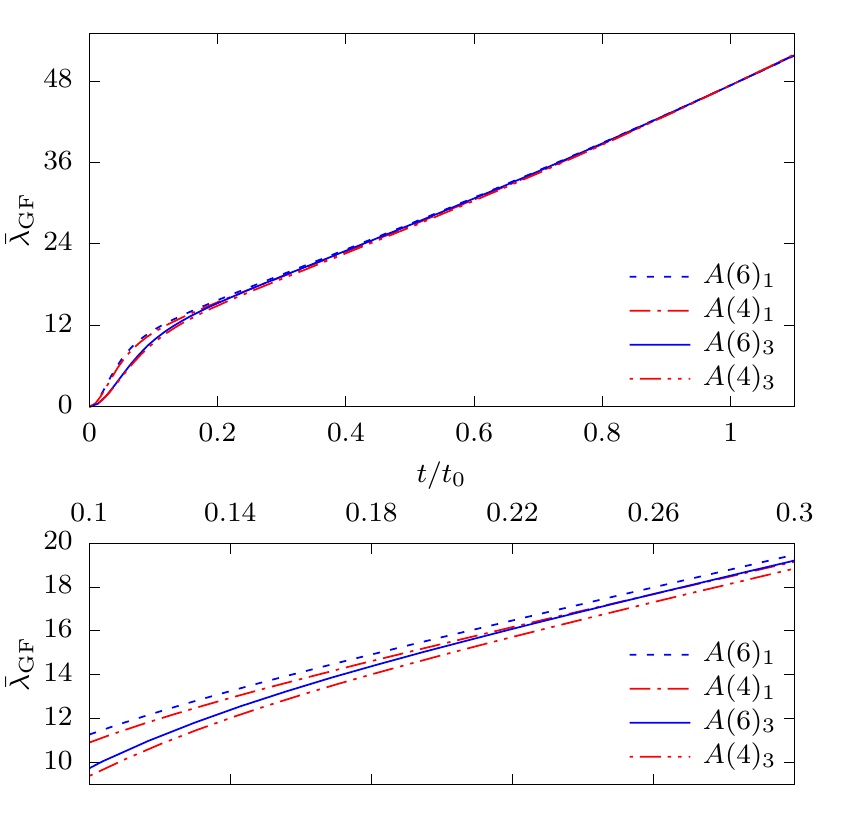}
\caption{$\GFcoup$ as a function of $t$ for several values of $\Nc$ and 
$a$ (see \Tab\ref{tab:fact_ensembles}). In the lower plot, we present a 
closer look at the small $t$ region.}
	\label{fig:lambdavst}
\end{figure}

In \Fig\ref{fig:lambdavst} we plot $\GFcoup$ as a function of $t$ for 
several gauge groups and different lattice spacings. Within the scale of 
the plot, the results are hard to distinguish for all gauge groups, 
which already shows the small size of the $\Nc$ dependent corrections. 
While at $t=t_0$ indepencence of $\Nc$ is enforced by \Eq\ref{eq:Gflow_coupling}
or equivalently
\begin{equation}
 \GFcoup(1/\sqrt{8t_0}) = 0.3 \times 16\pi^2\,,
 \label{eq:Gflow_couplingt0}
\end{equation}
the different $\Nc$ curves remain remarkably close when $t$ is a factor of 5 away from $t_0$.
At a 
closer look, corrections to $\Nc = \infty$ are present and the
data agrees very well with a polynomial in 
$1/\Nc^2$ as expected. We verified this by interpolating the data to 
several values of $t$ in a regular interval from $t=0.1 \, t_0$ to 
$t = 1.1 \, t_0$, and then taking the large $\Nc$ limit once at a fixed 
lattice spacing and once in the continuum. 

\begin{figure}
	\centering
	\includegraphics[width=\textwidth]{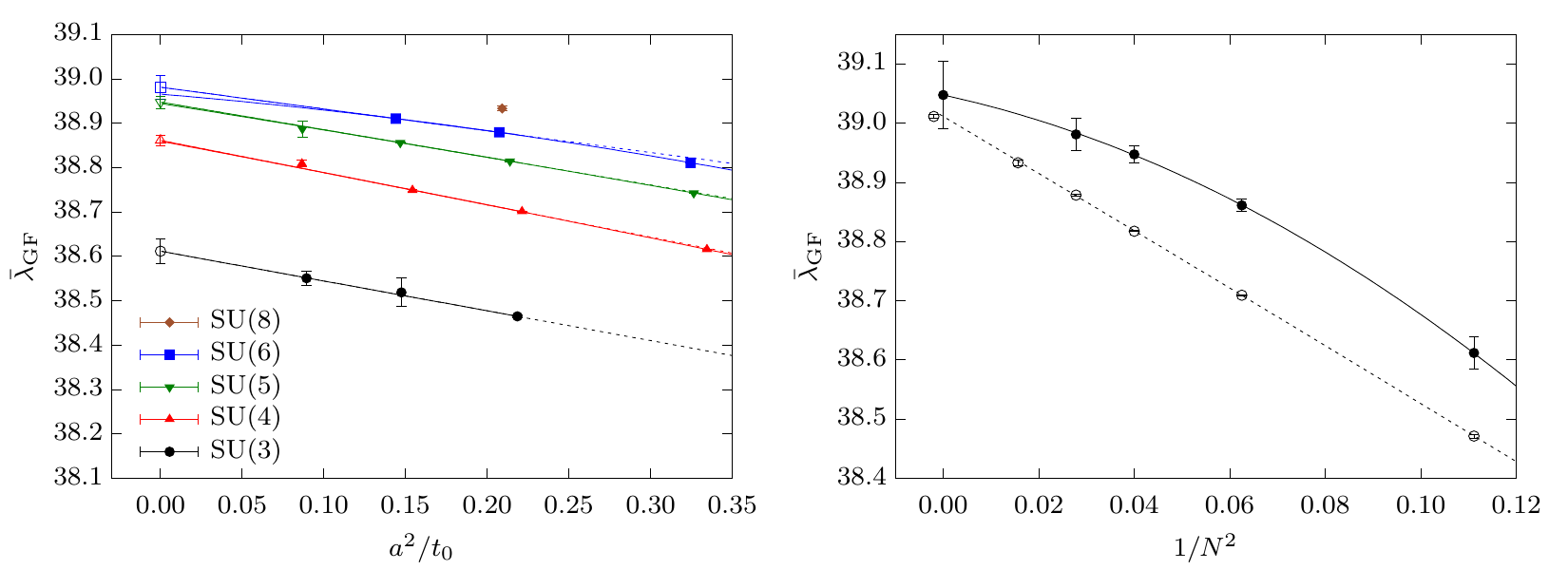}
	\caption{\textit{Left: }Continuum extrapolation of $\GFcoup$ 
		at $t/t_0 = 0.8$ for all the different gauge groups. 
		\textit{Right: } Large $\Nc$ extrapolations of $\GFcoup$
		in the continuum limit (solid line) and at finite lattice 
		spacing, $a^2/t_0=0.2091$, (dotted line). 
 The points at finite lattice 
	spacing have been slightly shifted for better legibility.
}
	\label{fig:lambdavsNt8}
\end{figure}

As can be observed in \Fig\ref{fig:lambdavst}, cut-off effects are 
large at small $t$. At $t= 0.1 \, t_0$ the relative difference between 
the results at the finest lattice spacing ($a \approx 0.05 \, \fm$) and 
at the coarsest ($a \approx 0.1 \, \fm$) one, is around $20\%$; while 
the errors in the measurements themselves is at the per-mill level. The 
situation is better at larger values of $t$, so let us first focus on 
values of $t/t_0 \geq 0.3$, where the relative size of cut-off effects is 
reduced tenfold, when compared to the case at $t/t_0=0.1$. In 
\Fig\ref{fig:lambdavsNt8} we 
show a plot of the continuum extrapolation of $\GFcoup$ at $t/t_0 = 0.8$
and the large $\Nc$ extrapolation both at finite lattice spacing and in 
the continuum. In order to be able to use the dataset at $\Nc =8$, in addition to 
the continuum limit extrapolations, 
we consider $a^2/t_0=0.2091$, the value on ensemble A(8)$_2$.
We then interpolated 
the results for 
all the other gauge groups to that lattice resolution.

On the left panel of 
\Fig\ref{fig:lambdavsNt8} we show the continuum limit extrapolations. 
The strategy chosen for the extrapolation is the following:
all continuum extrapolations are performed by linear
fits in $a^2/t_0$ to those data which satisfy  
$a^2/t_0 \leq 1/4$ (default fit). Such a restriction has been
well motivated in \Ref\cite{DallaBrida:2016kgh} for $\Nc=3$
and we find smaller discretization effects for larger $\Nc$.
As an estimate of the systematic uncertainty associated
with this choice, we perform a second fit linear in
$a^2$ with a data point at larger $a^2$; if the latter fit 
does not have a good $\chi^2$ we add an $a^4/t^2$ term to
the fit-function (control fit). 
If necessary, the error of the default fit is enlarged 
until it covers the full 1-$\sigma$ band of the control fit. The uncertainties of the continuum limit points
are usually dominated by the systematics which arises from 
different fits and which is \emph{not} necessarily 
independent for different $\Nc$.

All values of $\chidof$ are excellent except for \SU{4}, where
we obtain a value of $2.2$ and $2.7$ for the linear and quadratic fits 
respectively.  After performing the fits in $\ats$, on the right panel of 
\Fig\ref{fig:lambdavsNt8} we plot the large $\Nc$ extrapolations 
both in the continuum and at finite lattice spacing. As discussed, $\Nc =8$ is available only at finite lattice spacing, where 
in addition, errors are much smaller due to the fact that we performed an 
interpolation instead of the continuum limit 
extrapolation. The large $\Nc$ extrapolation uses the form

\begin{equation}
	Y(1/\Nc^2) = a_0 \left( 1 + \frac{a_1}{\Nc^2} + 
	\frac{a_2}{\Nc^4} \right) \, .	
	\label{eq:largeNfit}
\end{equation}

As seen in \Fig\ref{fig:lambdavsNt8}, the fit to the function $Y$ is 
excellent, with a $\chidof = 1.02$ 
at finite lattice spacing; for the continuum points we do not consider $\chi^2$ 
since the errors are 
strongly correlated due to the dominating systematic uncertainty
of the continuum extrapolations. Notice also that the results suggest that 
cut-off effects decrease with increasing $\Nc$. 

\begin{figure}
	\centering
	\includegraphics[width=\textwidth]{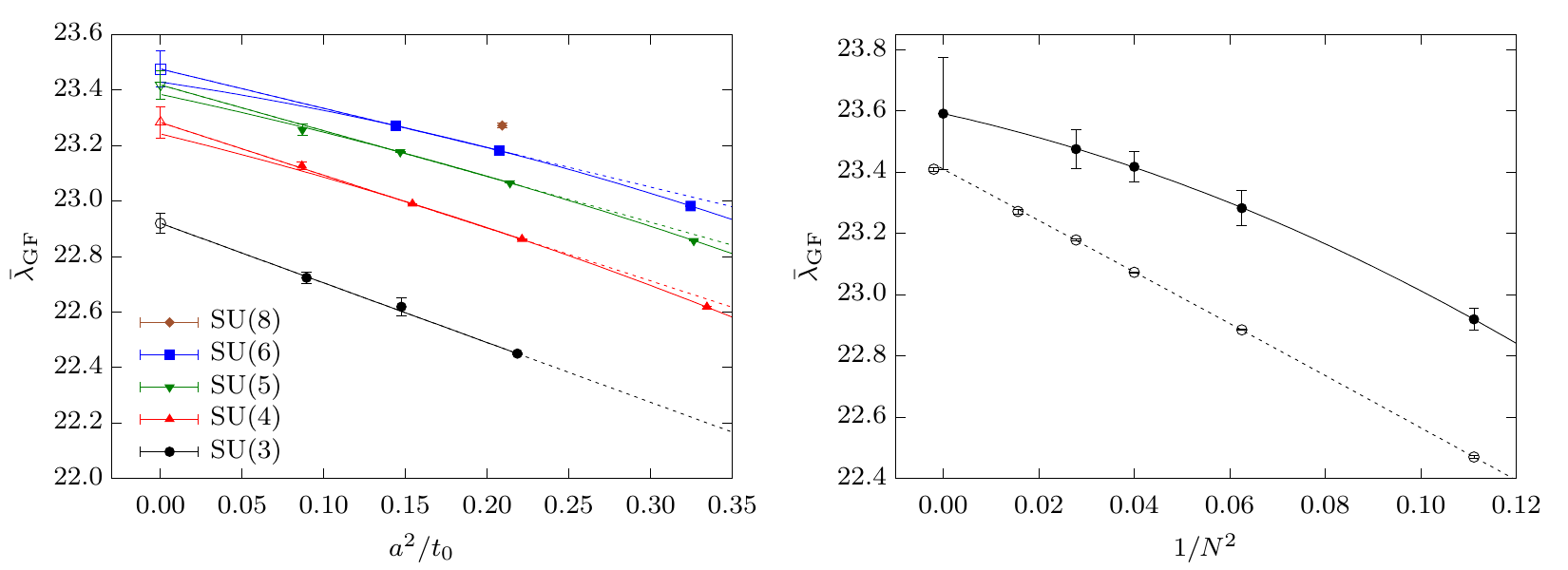}
	\caption{\textit{Left: }Continuum extrapolation of $\GFcoup$ 
		at $t/t_0 = 0.4$ for all the different gauge groups. 
		\textit{Right: } Large $\Nc$ extrapolations of $\GFcoup$
		in the continuum (solid line) and at finite lattice 
		spacing (dotted line). 
The points at finite lattice 
	spacing have been slightly shifted for better legibility.
}
	\label{fig:lambdavsNt4}
\end{figure}

As an example of results at smaller $t$, we
show our 
analysis at $t/t_0=0.4$  in 
\Fig\ref{fig:lambdavsNt4}. In this case, the magnitude of the cut-off 
effects is larger, but the same analysis as before can be carried out. 

As mentioned earlier, dealing with $\GFcoup$ at values 
of $t/t_0 \leq 0.3$ presents a bigger challenge, so one cannot reach 
the same level of accuracy as the results presented in this section. 
However, we have proceeded to do a similar analysis for such small values 
of $t$, including corrections of higher order in $\ats$. Details are found in \App\ref{sec:smallt}.

From the above analysis, we find that the large $\Nc$ dependence of $\GFcoup$
is in excellent agreement with the $1/\Nc^2$ scaling predicted by
the \tHooft perturbative expansion. Moreover, defining
\begin{equation}
	\eta(1/\Nc^2)=\Big\lvert \frac{Y(0) - Y(1/\Nc^2)}{Y(0)} \Big\rvert\, ,
	\label{eq:su3diff}
\end{equation}
we can determine the ``distance'' between \SU{3} and \SU{\infty}.  
In the continuum, at $t/t_0 = 0.8$ and $t/t_0=0.4$ we find 
$\eta(1/9) = 1.1\%$ and $2.8\%$ respectively.  Note also that the large 
$\Nc$ limit is taken at fixed $t_0$ and therefore $\eta\equiv 0$ at 
$t=t_0$ by definition. To account for this effect, we also fit 
$Y$ to $\GFcoup(1/\sqrt{8t}) - \GFcoup(1/\sqrt{8t_0})$ instead of $\GFcoup(1/\sqrt{8t})$, and 
define $\delta$ in a similar way to $\eta$. The results, together 
with those obtained for $\eta$ are displayed in \Tab\ref{tab:bParamW}.  
Let us remark that 
the individual errors in our measurements are below the per-mill 
level, so we can confidently quantify these percent level deviations 
between \SU{3} and \SU{\infty}.

The magnitude of 
the $1/\Nc^2$ corrections can be read off from
the coefficients 
$a_1$ and $a_2$ collected in \Tab\ref{tab:bParamW}, together with those of the 
smooth Wilson loops which we discuss next.

\begin{table}
	\centering
	\begin{tabular}{cccrrrccc}
		\toprule
obs.		& $c$     & fit & $a_0$	& $a_1$		& $a_2$    & $\chidof$& $\eta(1/9)$ & $\delta(1/9)$ \\
\toprule
$\GFcoup$ & $0.2$ & L & $15.916(6)$ & $-0.472(15)$ & $-0.05(12)$  & $0.79$ & $0.05$	& $0.03$\\
$\GFcoup$ & $0.4$ & L & $23.410(6)$ & $-0.3567(90)$  & $-0.043(70)$ & $1.07$ & $0.04$	& $0.04$ \\
$\GFcoup$ & $0.8$ & L & $39.011(3)$ & $-0.1233(30)$  & $-0.011(25)$ & $1.02$ & $0.014$	& $0.06$ \\
\midrule
$\GFcoup$ & $0.2$ & C & $16.2(5)$ & $-0.08(96)$ & $-1.9(65)$   & $0.44$ & $0.03$	& $0.02$ \\
$\GFcoup$ & $0.4$ & C & $23.6(2)$ & $-0.15(26)$  & $-1.0(18)$  & $0.01$ & $0.03$	& $0.03$ \\
$\GFcoup$ & $0.8$ & C & $39.05(6)$ & $-0.045(45)$ & $-0.50(32)$  & $0.02$ & $0.011$	& $0.05$ \\
\midrule
$W $ & $1/2$	     & L & $0.7760(7)$ & $0.355(33)$  & $-0.09(24)$ & $0.49$ & $0.04$	& $-$ \\
$W $ & $1  $	     & L & $0.6575(3)$ & $0.449(22)$  & $0.66(22)$  & $0.49$ & $0.06$   & $-$\\
$W $ & $9/4$	     & L & $0.4228(7)$ & $0.626(85)$  & $2.67(82)$  & $0.51$ & $0.10$   & $-$\\
\midrule
$W $ & $1/2$	     & C & $0.792(4)$  & $0.17(19)$   & $0.8(13)$   & $0.11$ & $0.03$	& $-$\\
$W $ & $1  $	     & C & $0.666(3)$  & $0.54(14)$   & $-0.5(10)$  & $0.01$ & $0.05$   & $-$\\
$W $ & $9/4$	     & C & $0.426(9)$  &  $1.27(69)$  & $-3.4(46)$  & $0.24$ & $0.10$   & $-$\\
\bottomrule
	\end{tabular}
	\caption{Parameters of the large $\Nc$ extrapolations, \eq(\ref{eq:largeNfit}), of 
	$\GFcoup(1/\sqrt{8ct_0})$ and $W(c)$ at finite lattice spacing (L) and in the continuum (C).}
	\label{tab:bParamW}
\end{table}

\subsubsection{Smooth Wilson loops}

\begin{figure}
	\centering
	\includegraphics[width=\textwidth]{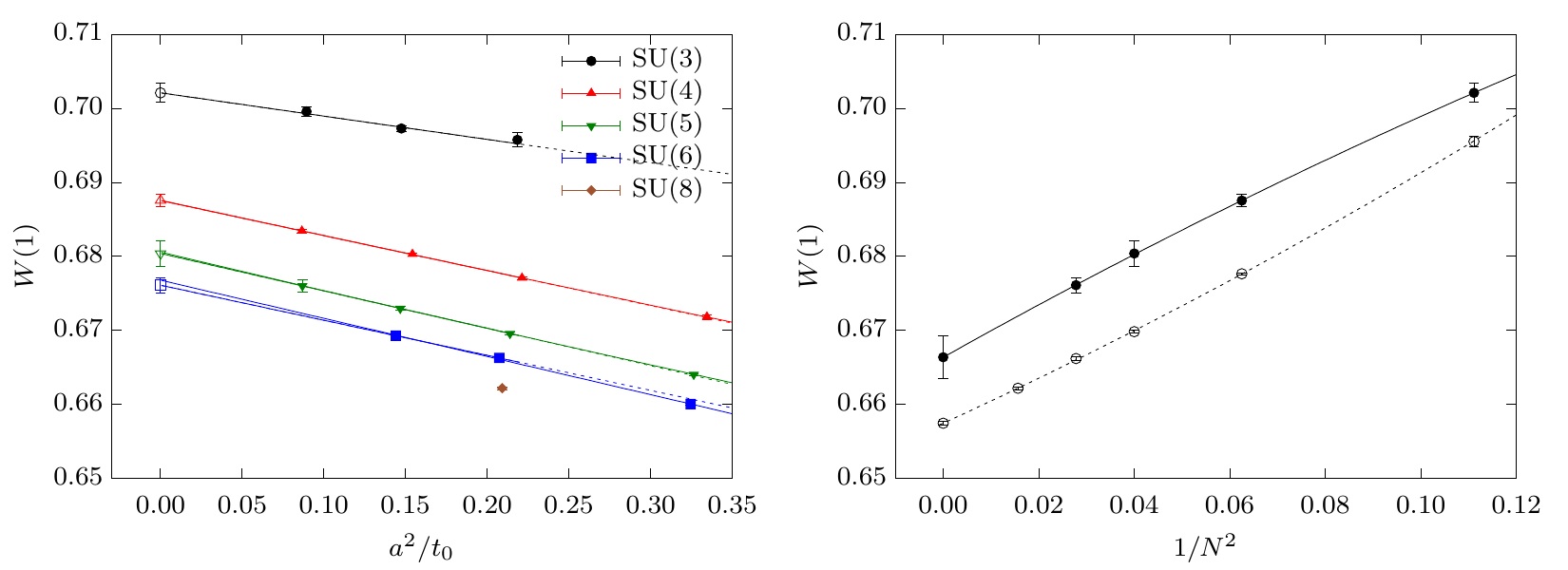}
	\caption{\textit{Left: }Continuum extrapolation of $W(1)$ for 
		all gauge groups. 
		\textit{Right: } Large $\Nc$ extrapolations of $W(1)$
		in the continuum (solid line) and at finite lattice 
		spacing (dotted line). There is an excellent agreement 
between the data and the expected scaling in powers of $1/\Nc^2$.
The points at finite lattice 
	spacing have been slightly shifted for better legibility.
}
	\label{fig:Wextrap}
\end{figure}

We have determined the smooth Wilson loops
at three different values of $c$, i.e. $c=1/2,1,9/4$. As in the case of 
$\GFcoup$, we are interested in the large $\Nc$ scaling at finite lattice
spacing and in the continuum. The strategy for the continuum limit fits 
is the same as for $\GFcoup$.
The fits for the loops at different 
$c$ are qualitatively similar, so in \Fig\ref{fig:Wextrap} we show the 
results at $c=1$ only. 

Once again, to quantify the magnitude of the finite $\Nc$ corrections, we
collect in 
\Tab\ref{tab:bParamW} the values of $a_1$ and $a_2$ from the 
fit to $Y$. We observe that the relative magnitude of them grow at larger
values of $c$ (or $t$ equivalently). Similarly, the deviation between 
\SU{3} and \SU{\infty} also grows up to a value of $\eta(1/9)=0.1$ when $c=9/4$. 
In all cases we find an excellent fit to $Y$ (the values of 
$\chidof$ are reported in \Tab\ref{tab:bParamW}).

\subsection{Factorization}

In order to verify the property of factorization from 
\Eq\eqref{eq:basic_fact}, we take the large $\Nc$ limit of the 
observables defined in \Sec\ref{sec:factobs}. The large $\Nc$ limits 
are taken in a similar way as described earlier, but we modify the 
parametrization of the large $\Nc$ fitting function for convenience, so 
that 

\begin{equation}
	Y(1/\Nc^2) = b_0 + \frac{b_1}{\Nc^2} + 
	\frac{b_2}{\Nc^4} \, .	
	\label{eq:largeNfit_fac}
\end{equation}

For the continuum limit extrapolations we use the same strategy as 
for $W$ and for $\GFcoup$, and in all cases, the data can be fitted very
well with a linear or quadratic polynomial in $\ats$. We 
also check for effects caused by variations of $L/\sqrt{8t}$ 
in all observables. As discussed in 
\App\ref{sec:app_finvol}, we find that $\HW$ at $c=1/2$ and 
$c=1$, are potentially affected by large effects. We tried 
to include them as a systematic error on the measurements, but 
this yields errors which are too large to be of interest as a test of 
factorization.
Hence, we present only results for $\HW$  at $c=9/4$ .

\begin{figure}
	\centering
	\includegraphics[width=\textwidth]{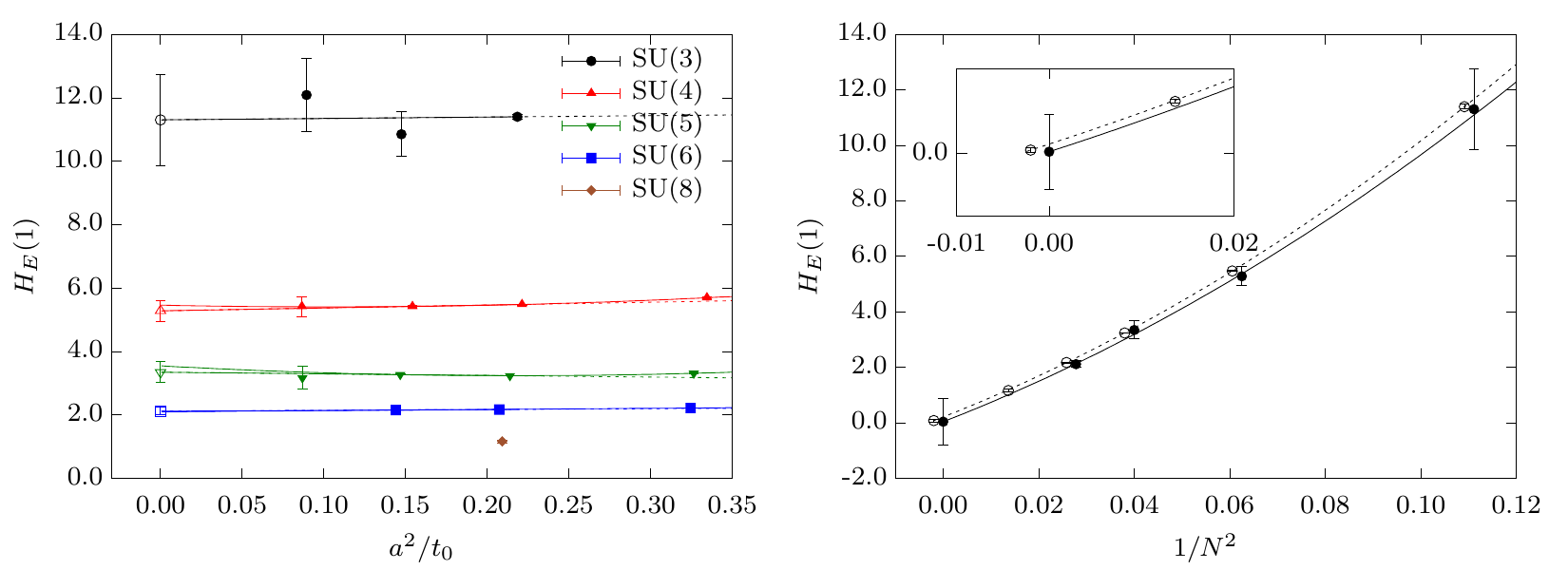}
	\caption{\textit{Left: }Continuum extrapolation of $\HE(1)$ 
		for all gauge groups. 
		\textit{Right: } Large $\Nc$ extrapolations of $\HE(1)$
		in the continuum (solid line) and at finite lattice 
		spacing (dotted line). The points at finite lattice 
	spacing have been slightly shifted for better legibility.}
	\label{fig:Fextrap}
\end{figure}

Let us first discuss our results for $\HE(1)$. On the right panel of 
\Fig\ref{fig:Fextrap} we show the large $\Nc$ fits both in the continuum 
and at a finite lattice spacing. The fits are excellent, which provides 
yet again confirmation of the scaling in powers of $1/\Nc^2$. It is worth 
mentioning that at finite lattice spacing, where results are very precise, 
we find that a quadratic fit 
in $1/\Nc^2$, excluding the \SU{3} point, extrapolates to \SU{3} within 
one standard deviation. In this sense, \SU{3} can be used as validation
of our fitting strategy. The values of the parameters of the 
fitting function $Y$ are displayed in \Tab\ref{tab:largeNparam}. At 
finite lattice spacing we include also the parameters from the fit 
excluding \SU{3}.

\begin{table}
	\centering
	\begin{tabular}{clrrrc}
		\toprule
obs.         & fit  & $b_0$      & $b_1$       & $b_2$       & $\chidof$  \\\toprule
$\HE(1)$   & L  & $0.078(44)$   & $66.3(19)$  & $319(19)$   & $0.30$ \\
$\HE(1)$   & L$_4$  & $0.069(75)$     & $66.8(35)$  & $314(43)$   & $0.59$ \\
$\HE(1)$   & C  & $0.04(84)$    & $67(39)$    & $293(373)$  & $0.34$ \\
$\HE(1)$   & L* & $0.0$         & $70.22(60)$ & $276(12)$   & $0.71$ \\
$\HE(1)$   & C* & $0.0$         & $70.3(98)$  & $237(200)$  & $0.27$ \\
\midrule
$\GW(1)$   & L  & $-0.00029(34)$ & $0.548(13)$ & $-0.69(12)$ & $0.02$ \\
$\GW(1)$   & L$_4$  & $-0.00017(69)$ & $0.540(44)$ & $-0.59(54)$ & $<0.01$ \\
$\GW(1)$   & C  & $-0.00003(36)$ & $0.49(13)$  & $-0.48(86)$  & $<0.01$ \\
$\GW(1)$   & L* & $0.0$ & $0.5376(58)$ & $-0.619(76)$  & $0.25$ \\
$\GW(1)$   & C* & $0.0$ & $0.485(28)$ & $-0.48(27)$ & $<0.01$ \\
\midrule
$\GW(1/2)$ & L  & $0.000045(373)$ & $0.167(17)$ & $-0.17(15)$ & $<0.01$ \\
$\GW(1/2)$ & L$_4$  & $-0.000008(646)$ & $0.170(42)$ & $-0.23(55)$ & $<0.01$ \\
$\GW(1/2)$ & C  & $-0.00049(247)$  & $0.168(82)$ & $-0.32(56)$   & $<0.01$ \\
$\GW(1/2)$ & L* & $0.0$         & $0.1686(70)$     & $-0.188(90)$ & $<0.01$ \\
$\GW(1/2)$ & C* & $0.0$         & $0.152(19)$    & $-0.22(18)$ & $0.02$ \\
\midrule
$\GW(9/4)$ & L  & $-0.00124(63)$ & $2.781(38)$ & $-3.95(43)$ & $0.13$ \\
$\GW(9/4)$ & L$_4$  & $-0.0010(10)$  & $2.760(74)$ & $-3.6(10)$  & $0.15$ \\
$\GW(9/4)$ & C  & $0.0014(99)$   & $2.43(40)$  & $-1.8(29)$  & $0.13$ \\
$\GW(9/4)$ & L*  & $0.0$         & $2.711(14)$ & $-3.23(24)$ & $1.39$ \\
$\GW(9/4)$ & C*  & $0.0$         & $2.485(73)$ & $-2.22(81)$ & $0.07$ \\
\midrule
$\HW(9/4)$ & L  & $-0.24(14)$	  & $120(9)$    & $-129(84)$   & $0.39$ \\
$\HW(9/4)$ & C  & $-3.3(25)$    & $226(88)$   & $-1063(618)$ & $0.26$ \\
$\HW(9/4)$ & L*  & $0.0$ 	  & $106(3)$    & $-7(40)$     & $1.17$ \\
$\HW(9/4)$ & C*  & $0.0$ 	  & $112(17)$   & $-290(186)$  & $0.99$ \\
\bottomrule
	\end{tabular}
	\caption{Parameters of the large $\Nc$ extrapolations of $\HE$,
	$\GW$ and $\HW$. We present the results for three different 
	cases, L: at finite lattice spacing, L$_4$: at 
	finite lattice spacing excluding \SU{3}, and C: 
in the continuum. Additionally, we fit the data to a function with $b_0 =0$, 
so that factorization is imposed at finite lattice spacing L* and in the 
continuum C*. In this case, the value of $\chidof$ 
validates this hypothesis.}
	\label{tab:largeNparam}
\end{table}

Concerning the $\Nc\to\infty$ limit itself, the extrapolated value is within
two standard deviations from 
zero in the worst case. Notice that at finite lattice spacing, the 
errors in the extrapolation are two orders of magnitude smaller than the 
value of $\HE(1)$ at $\Nc =3$. To further validate factorization, an 
additional fit is performed for which $b_0 =0$ is fixed, and only $b_1$ 
and $b_2$ are fitted to the data. This enforces factorization, so the 
value of $\chidof$ from the fit can be used to asses the validity of the 
assumption ($L^*,C^*$ in \Tab\ref{tab:largeNparam}). 
To summarize, for $\HE(1)$ we find 
excellent agreement with factorization in the continuum, and a deviation 
compatible with two standard deviations in the worst case at finite 
lattice spacing, still statistically consistent with 
factorization.

\begin{figure}
	\centering
	\includegraphics[width=\textwidth]{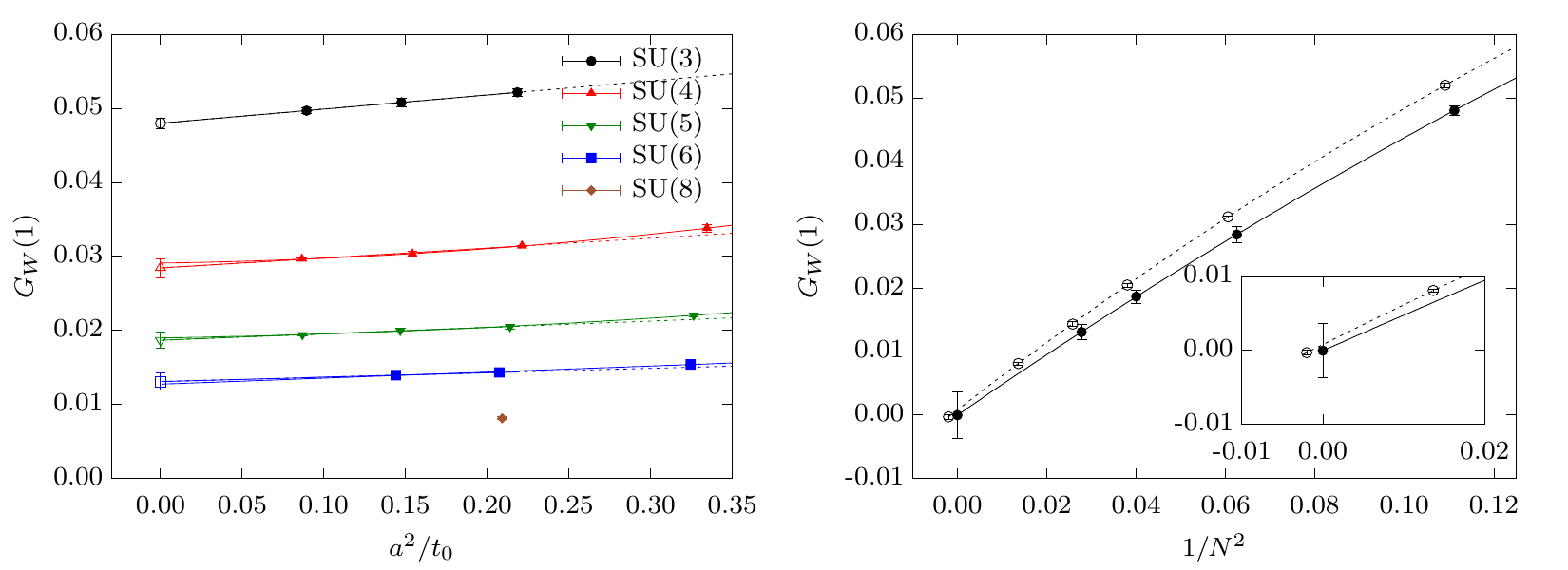}
	\caption{\textit{Left: }Continuum extrapolation of $G_W(1)$ for
		all gauge groups. 
		\textit{Right: } Large $\Nc$ extrapolations of $G_W(1)$
		in the continuum (solid line) and at finite lattice 
		spacing (dotted line). The points at finite lattice 
	spacing have been slightly shifted for better legibility.}
	\label{fig:Gextrap}
\end{figure}

We now turn to the smooth Wilson loops. In 
\Fig\ref{fig:Gextrap} we display the results of the continuum and large 
$\Nc$ fits for $\GW(1)$. The parameters of the extrapolations at the 
three values of $c$ are displayed in \Tab\ref{tab:largeNparam}. Also in 
this case, we find that \SU{3} can be used as a validation point, and if 
it is excluded from the fit, it agrees with the extrapolating function 
within two standard deviations at $c=1$, and within one standard 
deviation at the remaining values of $c$.

For the fits with factorization enforced by fixing $b_0 = 0$, the values of $\chidof$ 
are also excellent. These values, 
together with those of $b_0$ reported in \Tab\ref{tab:largeNparam}, give
us confidence on the validity of factorization. Notice 
that the errors at large $\Nc$ are at least one order of magnitude 
smaller than the value at \SU{3} itself. 
Concerning the finite $\Nc$ corrections, comparing the loops at 
different values of $c$, we observe that those 
at large $c$ are characterized by large coefficients in front of the 
$1/\Nc^2$ and $1/\Nc^4$ correction terms.  

\begin{figure}
	\centering
	\includegraphics[width=\textwidth]{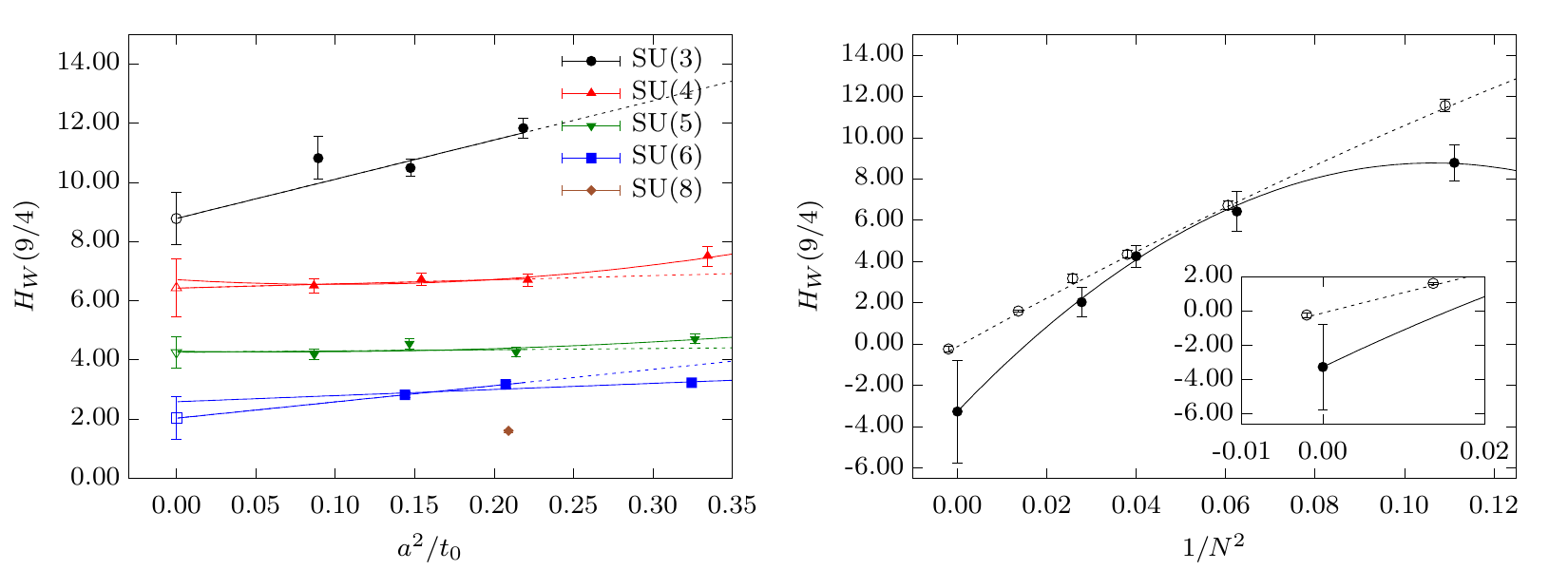}
	\caption{\textit{Left: }Continuum extrapolation of $\HW(9/4)$ 
		for all gauge groups. 
	\textit{Right: } Large $\Nc$ extrapolations of $\HW(9/4)$
		in the continuum (solid line) and at finite lattice 
		spacing (dotted line). The points at finite lattice 
	spacing have been slightly shifted for better legibility.}
	\label{fig:Hextrap}
\end{figure}

Yet another 
interesting question is whether loops at fixed $t$ but different $R$ have 
different finite $\Nc$ corrections. We explore this issue at the end of 
the section. Let us first look at $\HW$ in 
\Fig\ref{fig:Hextrap}.  The $\Nc\to\infty$ limits are less than two standard deviations 
away from zero.  Inspecting the continuum limit fits, we observe that 
had we taken the three-point extrapolation for $\Nc=6$ as our central result, the central value would have been close to the upper end of the error bar 
in \Fig\ref{fig:Hextrap} and the $1/\Nc^2$ extrapolation in full agreement
with factorization. In other words, one should not look too much at 
the central value but at the full range of the error, as always.

\subsubsection{Loop size dependence}

\begin{table}
	\centering
	\begin{tabular}{ccrrrrr}
		\toprule
$\xi$  & $a_0$        & $a_1$      & $a_2$       & $b_0$          & $b_1$	 & $b_2$  \\\toprule
$1.00$ & $0.7760(7)$  & $0.36(3)$  & $-0.09(24)$ & $0.00005(37)$  &  $0.167(17)$ & $-0.17(15)$ \\
$1.25$ & $0.5956(7)$  & $0.68(4)$  & $0.40(34)$  & $0.0003(13)$   &  $0.752(55)$ & $-1.13(44)$ \\
$1.50$ & $0.4087(5)$  & $1.16(5)$  & $1.34(48)$  & $0.0005(31)$   &  $2.62(12)$  & $-5.05(92)$ \\
$1.75$ & $0.2512(4)$  & $1.78(7)$  & $3.16(71)$  & $0.0053(65)$   &  $8.06(22)$  & $-18.7(17)$ \\
$2.00$ & $0.1390(3)$  & $2.48(13)$ & $6.7(12)$   & $0.030(13)$    &  $26.12(50)$ & $-81.0(45)$ \\
\bottomrule
	\end{tabular}
	\caption{Parameters of the large $\Nc$ extrapolation of $\hW$ 
		and $\hGW$ as a function of $\xi$.}
	\label{tab:Rcparam}
\end{table}

Finally, let us explore how the finite $\Nc$ corrections to 
factorization change when the size of the loop is increased at a fixed 
value of the smoothing parameter $t$. For a given value of $t$, we 
consider square loops of size $R(\xi) = \xi \sqrt{8t}$. Given the 
finite size of the lattices, we use the loops measured at $c=1/2$ 
($t=t_0/2$), so that we can consider larger values of $\xi$. 
Thus, at a fixed value of $t=t_0/2$, we define $\hW$ and $\hGW$ in a 
similar way as $W$ and $\GW$, but in this case, as a function
of $\xi$ instead of $c$. In addition to the already presented results
at $\xi=1$, we also measured $\hW$ and $\hGW$ at 
$\xi=1.25,\, 1.5, \, 1.75$ and $2$. The coefficients 
obtained for the large $\Nc$ fits at finite lattice 
spacing are displayed in \Tab\ref{tab:Rcparam} as a function of $\xi$.

\begin{figure}
	\centering
	\includegraphics[width=0.48\textwidth]{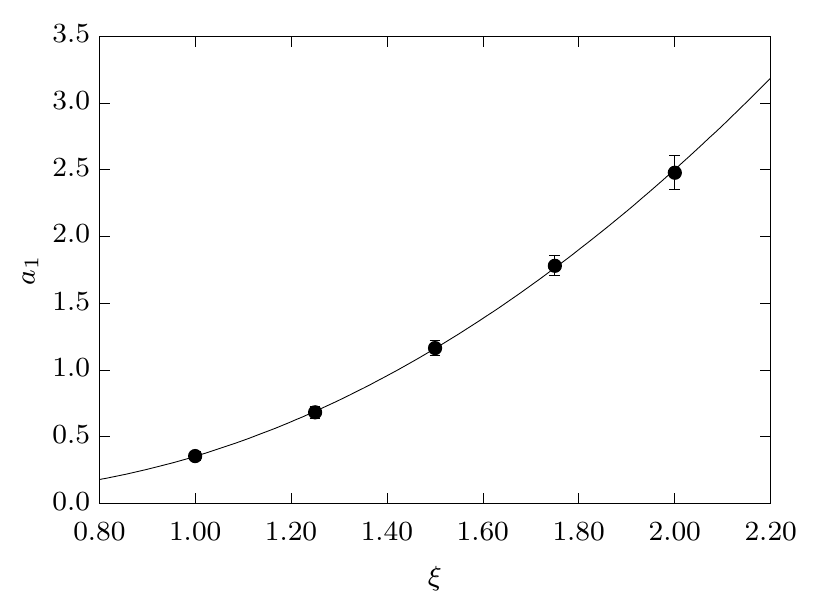}
	\includegraphics[width=0.48\textwidth]{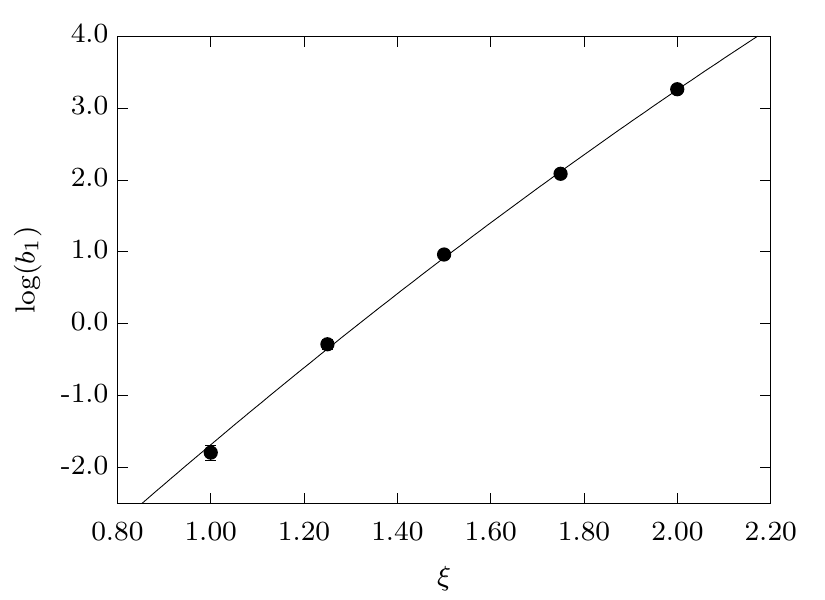}
	\caption{Plot of the parameters $a_1$ and $\log(b_1)$ as a function 
	of $\xi$. The interpolating function is a quadratic 
function in $\xi$ in both cases.}\label{fig:Rcvsw}
\end{figure}

We observe that in the case 
of the loops themselves, the coefficients of the $1/\Nc^2$ expansion do 
not change significantly with $\xi$, while those of $\hGW$ grow 
rapidly for larger loops. In fact, they grow exponentially fast as 
shown in \Fig\ref{fig:Rcvsw}. At finite $\Nc$, larger loops are
much further away from $\Nc \to \infty$ than smaller loops.

\section{Conclusions}

We have taken the large $\Nc$ limit of a few observable
of \SU{\Nc} pure gauge theories numerically defining all
dimensionfull quantities in units of $t_0$. This means that
we held $t_0$, or equivalently the coupling 
$\GFcoup$, \eq\eqref{eq:Gflow_coupling}, at a low energy fixed in defining the approach to the  limit. As explained in \sect\ref{sec:limit},
the precise magnitude of $1/\Nc^2$ corrections do depend on
this choice. For each quantity, the continuum limit was taken 
before the large $\Nc$ limit, but
we have also investigated large $\Nc$ scaling at finite lattice
spacing, defined by $a^2/t_0$=constant. 

In both cases we find that finite $\Nc$ observables are very well
and very precisely described by a leading order term and 
corrections $\sim 1/\Nc^2$ and $\sim 1/\Nc^4$. We recall for example Figure~\ref{fig:lambdavsNt4} where the excellent precision, in
particular at finite $a$, is visible. 
In the same way, factorization has been confirmed very precisely.
Of course, a numerical computation cannot substitute a mathematical proof, but our results make it very implausible that 
anything goes wrong with the large $\Nc$ limit in general,
or factorization in particular.

However, the {\em magnitude of corrections} to the large 
$\Nc$ limit is more complex. We found a strong
dependence on the physical size of the observables.
For example, we considered $R\times R$ Wilson loops 
smoothed with a smoothing radius of size 
again $\sqrt{8t}=R$. 
Table \ref{tab:bParamW} shows the deviation, $\eta$, of SU(3)
from SU($\infty$) of these smooth loops to increase
from 3\% at a loop-size of $r=0.2 \, \fm$  to  10\% at $R=1 \, \fm$.

When we increase the loop size $R$ at fixed smoothing radius
$\sqrt{8t}=0.23 \, \fm$ from 
$R=0.23 \, \fm$  to $R=0.5 \, \fm$, the corrections
$\eta(1/9) \approx a_1/9$ (with $a_1$ from Table \ref{tab:Rcparam}
or Figure \ref{fig:Rcvsw}) grow from 4\% to more than 30\%.
The growth with $R$ of the finite $\Nc$ corrections to factorization   
is even more dramatic as seen on the right panel in
Figure \ref{fig:Rcvsw}. These large corrections may also contribute 
to the fact that one has to go to very large  $\Nc$
to approach the large $\Nc$ limit in the 1-point model \cite{Gonzalez-Arroyo:2014dua,Gonzalez-Arroyo:2015bya}. Of course, the dominating effect
is expected to be that the color degrees of freedom provide the 
effective size of the system in that model.

In summary, large $\Nc$ scaling is confirmed with high precision,
but corrections to large distance observables can be 
substantial. One thus has to be
careful when deriving quantitative information from large 
$\Nc$ considerations in gauge theories.

\begin{acknowledgement}
We would like to thank M. C\`e, L. Giusti and S. Schaefer 
for sharing part of the generation of the gauge configurations \cite{Ce:2016awn}
and M. Koren for useful discussions. We are grateful to A. Gonzalez-Arroyo
and U. Wolff for their valuable comments on our first manuscript.
Our simulations were performed at the ZIB computer center with the 
computer resources granted by The North-German 
Supercomputing Alliance (HLRN). M.G.V acknowledges the 
support from the Research Training Group GRK1504/2 ``Mass, Spectrum, Symmetry'' 
founded by the German Research Foundation (DFG). 
\end{acknowledgement}

\begin{appendices}

\section{Volume dependence}\label{sec:app_finvol}

\begin{figure}
	\centering
	\includegraphics[width=\textwidth]{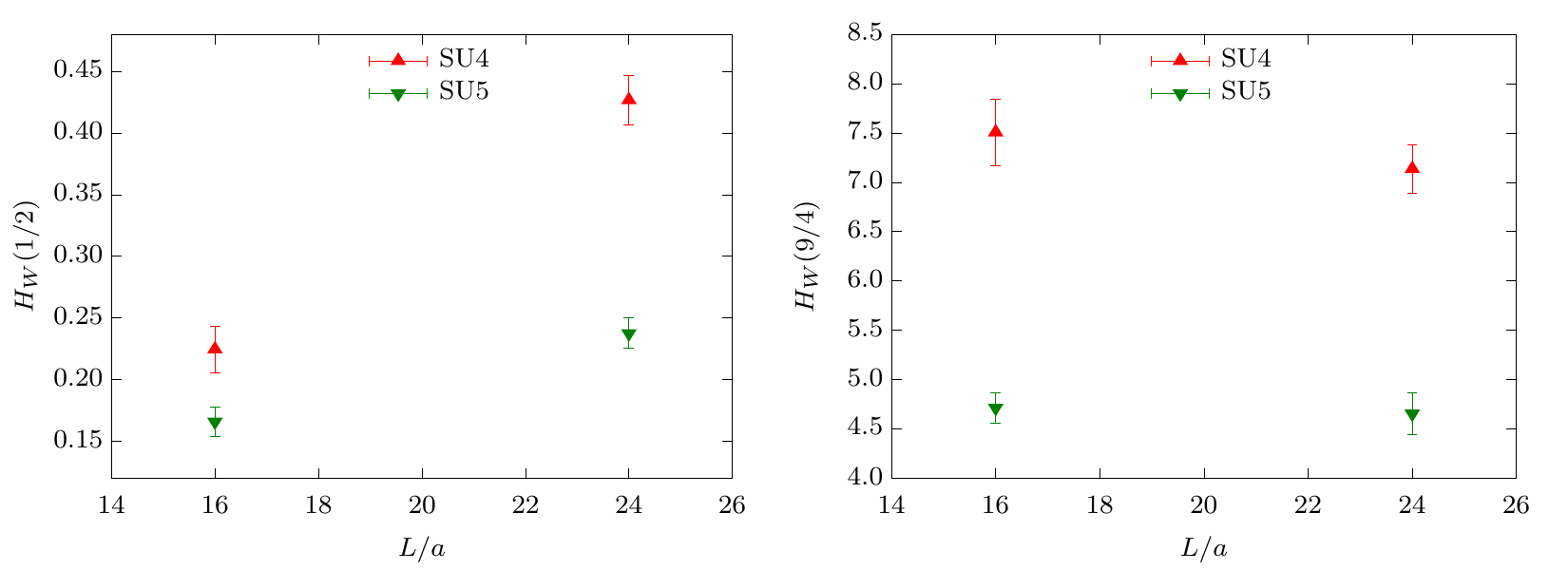}
	\caption{Finite volume checks for $\HW$. On the left at $c=1/2$,
		and on the right at $c=9/4$. Notice that at the smaller 
	$c$ the finite volume corrections are very significant.
}
	\label{fig:fvolHW}
\end{figure}

In order to understand whether $L$ is kept sufficiently
close to a fixed value, we 
have performed two additional simulations at the coarsest lattice 
spacing for \SU{4} and \SU{5}. The parameters for these simulations are 
the same as for $A(4)_1$ and $A(5)_1$ respectively, with the difference
that the lattice sizes have been increased to $24^3\times 96$. In 
physical units this corresponds to $L \approx 2.4\, \fm $. Although all 
the ensembles in \Tab\ref{tab:fact_ensembles} have approximately the 
same lattice size, we find that in the case of $\HW(c)$, at $c=1/2$ and 
$c=1$, the small variations in the volume induces an uncertainty which may
be relevant. For the rest
of observables, volume effects are within the statistical 
uncertainty. To showcase this, in \Fig\ref{fig:fvolHW} we display a plot of
$\HW(c)$ for $c=1/2$ and $c=9/4$. Clearly, at the smaller $c$, 
volume effects are much larger. Both at $c=1/2$ and $c=1$ we find that 
including the volume effects as a systematic correction is difficult
with just two points in $L$. Attempting it in a conservative manner
produces errors which are too large to check for the large $\Nc$ scaling
at a similar precision as for the rest of observables, so we include 
only $\HW(9/4)$ in the analysis. We note that we do not have an
explanation why small $c$ appears to be more difficult than a 
large one.

\section{$\GFcoup$ at small $t$}\label{sec:smallt}

\begin{figure}
	\centering
	\includegraphics[width=\textwidth]{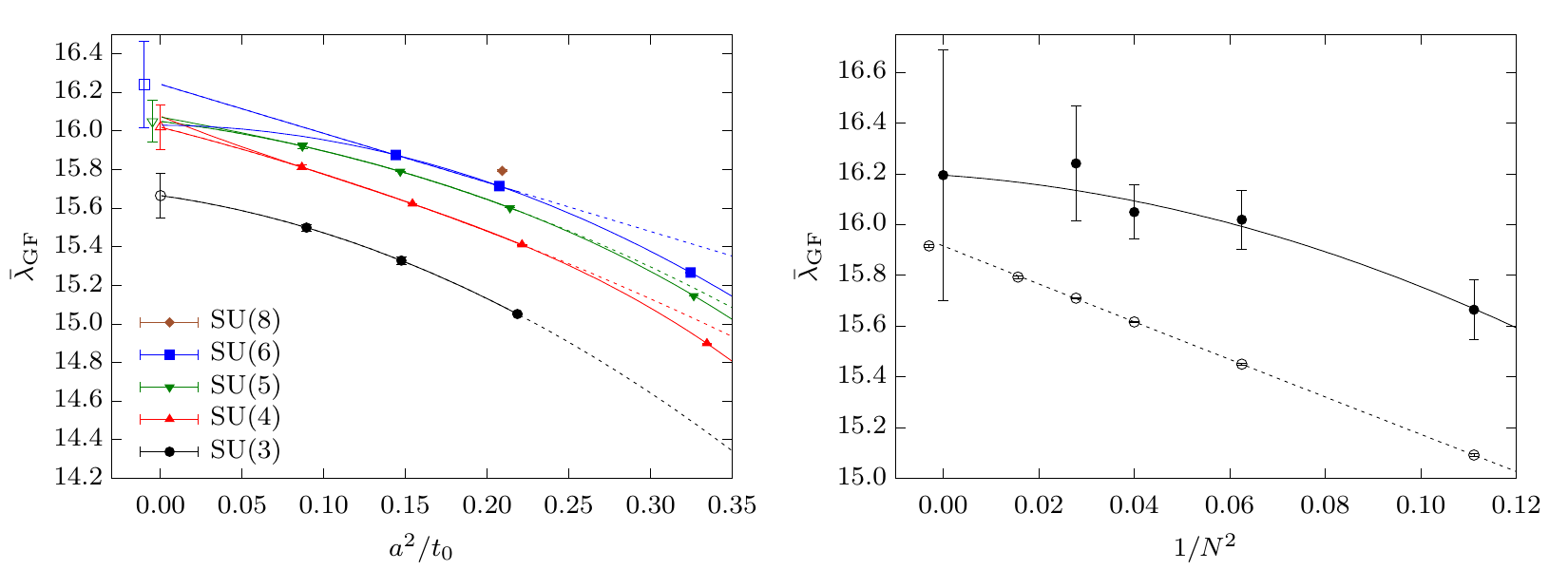}
	\caption{\textit{Left: }Continuum extrapolation of $\GFcoup$ 
		at $t/t_0 = 0.2$ for all the different gauge groups. 
		\textit{Right: } Large $\Nc$ extrapolations of $\GFcoup$
		in the continuum (solid line) and at finite lattice 
		spacing (dotted line). There is an excellent agreement 
between the data and the expected scaling in powers of $1/\Nc^2$.
The points at finite lattice spacing have been slightly shifted for 
better legibility.}
	\label{fig:lambdavsNt2}
\end{figure}

As stressed in \Sec\ref{sec:resgfcoup}, the continuum extrapolations 
become more difficult at smaller values of $t$. Due to the large 
cut-off effects, we find that a linear extrapolation in $\ats$ does not 
parametrize the data adequately, even using only the finest lattice 
points. For that reason, we include in the fits the $\rmO(a^4/t_0^2)$ and 
the $\rmO(a^6/t_0^3)$ corrections. The fit strategy is similar to the one 
used for the rest observables, except that higher degree polynomials are used.
Briefly, the central point is obtained by performing a quadratic fit in 
$a^2/t_0$ using those data for which $a^2/t_0 \leq 1/4$. Then, a second fit 
is performed, either using a quadratic function, or a cubic one if the 
value of $\chidof$ of the first one is too large. Finally, the error is chosen 
so that it covers the full 1-$\sigma$ band of both fits. We show our results at 
$t/t_0 = 0.2$ in 
\Fig\ref{fig:lambdavsNt2}. As expected, the errors in our continuum 
extrapolations are larger than those obtained at larger values of 
$t$, but within the errors, the large $\Nc$ extrapolation is 
perfectly consistent with a polynomial in $1/\Nc^2$. At finite lattice 
spacing, the large $\Nc$ extrapolation is cleaner, and once again it 
shows and excellent agreement with the \tHooft expansion.

\end{appendices}

\bibliographystyle{JHEP}
\usebiblio{references}


\providecommand{\href}[2]{#2}\begingroup\raggedright\endgroup

\end{document}